\documentclass[twocolumn]{aastex701}
\usepackage[utf8]{inputenc}   

\usepackage{gensymb}

\newcommand{\hb}{H$\beta$}
\newcommand{\ha}{H$\alpha$}

\newcommand{\oiii}{[\ion{O}{3}]}
\newcommand{\nii}{[\ion{N}{2}]}
\newcommand{\sii}{[\ion{S}{2}]}

\begin{document}

\title{Clumpy Disk, Interloper, or Merger? Nature of a Distant Galaxy Pair at 5 kpc Projected Separation}

\correspondingauthor{Hajar Aziz}
\email{haziz005@ucr.edu}

\author[0000-0003-1892-3121]{Hajar Aziz}
\affiliation{Department of Physics and Astronomy, 4129 Frederick Reines Hall, University of California, Irvine, CA 92697, USA}
\affiliation{Department of Physics and Astronomy, University of California, Riverside, CA 92521, USA}
\email{haziz005@ucr.edu}

\author[0000-0002-1912-0024]{Vivian U}
\affiliation{IPAC, California Institute of Technology, 1200 E. California Blvd., Pasadena, CA 91125, USA}
\affiliation{Department of Physics and Astronomy, 4129 Frederick Reines Hall, University of California, Irvine, CA 92697, USA}
\email{vivianu@ipac.caltech.edu}

\author[0000-0001-7578-2412]{Archana Aravindan}
\affiliation{Department of Astronomy, The University of Texas at Austin, Austin, TX 78712, USA}
\affiliation{Cosmic Frontier Center, The University of Texas at Austin, Austin, TX 78712, USA}
\affiliation{Department of Physics and Astronomy, University of California, Riverside, CA 92521, USA}
\email{archana.aravindan@email.ucr.edu}

\author[0000-0002-0164-8795]{Raymond P. Remigio}
\affiliation{Department of Physics and Astronomy, 4129 Frederick Reines Hall, University of California, Irvine, CA 92697, USA}
\email{remigior@uci.edu}

\author[0000-0003-4693-6157]{Gabriela Canalizo}
\affiliation{Department of Physics and Astronomy, University of California, Riverside, CA 92521, USA}
\email{gabyc@ucr.edu}

\author[0000-0002-6650-3757]{Justin A. Kader}
\affiliation{IPAC, California Institute of Technology, 1200 E. California Blvd., Pasadena, CA 91125, USA}
\email{jkader@ipac.caltech.edu}

\author[0000-0002-6570-9446]{Marina Bianchin}
\affiliation{Instituto de Astrof\' isica de Canarias, Calle V\'ia L\'actea, s/n, E-38205, La Laguna, Tenerife, Spain}
\affiliation{Departamento de Astrof\'isica, Universidad de La Laguna, E-38206, La Laguna, Tenerife, Spain
}
\affiliation{Department of Physics and Astronomy, 4129 Frederick Reines Hall, University of California, Irvine, CA 92697, USA}
\email{marina.bianchin@iac.es}

\author[0000-0002-3139-3041]{Yiqing Song}
\affiliation{European Southern Observatory, Alonso de C{\'o}rdova, 3107, Vitacura, Santiago, 763-0355, Chile}
\affiliation{Joint ALMA Observatory, Alonso de C{\'o}rdova, 3107, Vitacura, Santiago, 763-0355, Chile}
\affiliation{Max-Planck-Institut f{\"u}r Radioastronomie, Auf dem H{\"u}gel 69, 53121, Bonn, Germany}
\email{Yiqing.Song@eso.org}

\author[0000-0002-1158-6372]{Tianmu Gao}
\affiliation{Research School of Astronomy and Astrophysics, Australian National University, Weston Creek, ACT 2611, Australia
}
\affiliation{ARC Centre of Excellence for All Sky Astrophysics in 3 Dimensions (ASTRO 3D); Australia}
\email{tianmu.gao@anu.edu.au}

\author[0000-0002-5807-5078]{Jeff Rich}
\affiliation{The Observatories of the Carnegie Institution for Science, 813 Santa Barbara Street, Pasadena, CA 91101, USA}
\email{jrich@carnegiescience.edu}

\author[0000-0003-2064-4105]
{Rosalie McGurk}
\affiliation{W. M. Keck Observatory, 65-1120 Mamalahoa Hwy. Kamuela, HI 96743, USA}
\email{rmcgurk@keck.hawaii.edu}

\begin{abstract}
We present the morphological, photometric, and spectroscopic properties of a $z \sim 1 $ galaxy, ``lil gal'', serendipitously detected in \textit{JWST} Mid Infrared Instrument (MIRI) images of nearby galaxy  VV 340. In the MIRI F560W and F770W images, we identify what appears to be a spiral galaxy with a central bulge. However, in the F1500W image, a second peak appears $\sim 0.\arcsec7$ northwest (NW) from the central bulge, calling into question the nature of this source as a clumpy disk, a high-redshift interloper, or a galaxy merger. Multi-band analyses of the three MIRI and four \textit{Hubble Space Telescope (HST)} images suggest a photometric redshift of \(\sim 0.92\). Spectroscopic analyses of data from the Keck Near-Infrared Echellette Spectrometer (NIRES) reveal two sets of \nii~and \ha~emission lines corresponding to the two observed sources. A redshift of \(z = 0.9248\) is identified for the NW companion. Fainter emission lines are identified from the underlying galaxy at \(z = 0.9225\), suggesting a merging galaxy pair at a projected separation of \(\sim 5\) kpc. From the emission line ratios from Keck NIRES and Keck Cosmic Web Imager (KCWI) spectra, we classify the system as hosting regions of active star formation, likely attributed to merger-induced starburst activity. The results demonstrate the necessity of resolved, spectroscopic follow-up analyses of galaxies found in deep \textit{JWST} images to disentangle the role of galaxy mergers from clumpy disk galaxies at $z \sim 1$ to cosmic noon and beyond.
\end{abstract}

\section{Introduction} \label{sec:intro}
Studying galaxy formation and evolution provides crucial insights into the fundamental processes that have shaped the universe. In particular, the role of galaxy mergers in triggering star formation, active galactic nuclei (AGN) activity, and overall structural changes remains a topic of ongoing investigation \citep{lambas_galaxy_2012, ellison_galaxy_2013, 2024A&A...691A..82C}.

Over recent decades, considerable work from \textit{HST} studies have focused on galaxy morphology to understand the role of mergers at intermediate redshifts. Earlier studies of the Medium Deep Survey found that $\sim40\%$ of irregular galaxies showed evidence of interactions or multiple core structure, with $\sim60\%$ of the multiple-core irregulars showing close companions \citep{driver_contribution_1995}. \cite{conselice_direct_2003} formalized this approach by applying the concentration, asymmetry, clumpiness (CAS) morphological system to galaxies detected out to $z \sim 3$ in the Wide Field Planetary Camera 2 (WFPC2) and the Near Infrared Camera and Multi-Object Spectrometer (NICMOS) in the Hubble Deep Field North, using asymmetry as a key criterion for identifying mergers. However, \cite{de_propris_millennium_2007} details limitations of the CAS selection system of image asymmetry, reporting false positives in merger classifications due to contamination from nearby stars. Later studies by \cite{duncan_observational_2019}, from the Cosmic Assembly Near-infrared Deep Extragalactic Legacy Survey (CANDELS), highlighted the lack of observational constraints on the merger history of galaxies at $z > 2$ and reported an increase in merger rates up to $z \sim 6$, which differs from previous work in which the merger rate appeared to decline at higher redshifts \citep{conselice_structures_2009}. Collectively, these \textit{HST} studies demonstrate how the identification of merging systems has relied on irregular and asymmetric morphologies. However, because \textit{HST} observations---particularly of higher redshift sources---are strongly affected by dust extinction, the observed morphologies may be biased by dust, highlighting the need for more detailed follow-up studies to better understand mergers and their role in galaxy evolution.

Infrared studies conducted with the \textit{Spitzer Space Telescope} further advanced our knowledge by probing the physical properties of galaxies at \(z \sim 0.3-3.0\), distinguishing populations dominated by starburst or AGN activities \citep{floch_infrared_2005, geach_spitzer_2007, sajina_spitzer-_2012}. \textit{Spitzer} observations also highlighted discrepancies in earlier findings, challenging the previously assumed dominance of irregular and peculiar morphologies at $z \sim 0.5$ by reporting the prevalence of disk galaxies within this epoch, albeit with significant asymmetries indicative of mergers \citep{shi_morphology_2006,lotz_evolution_2008}. However, \textit{Spitzer’s} limited spatial resolution—0.85 m primary mirror compared to \textit{HST’s} 2.4 m—posed challenges for detailed morphological analyses, underscoring the need for higher-resolution infrared observations.

The difficulty of merger identification is particularly pronounced at $z \sim 1$, an epoch past the peak of star formation and merger activity. \cite{kartaltepe_multiwavelength_2010} conducted visual morphological classifications of Spitzer 70 \(\mu\)m selected galaxies in the COSMOS field \added{\citep{2007ApJS..172...38S}} using \textit{HST} ACS imaging and reported that at $0.5 < z < 1.0$, $\sim26\%–43\%$ of LIRGs are major mergers, with the fraction rising to $\sim47\%–88\%$ for ULIRGs. \added{Similarly, \cite{2007A&A...468...33E} found that nearly half of LIRGs at $z \sim 1$ present spiral morphologies in HST imaging, with only $\sim31\%$ exhibiting clear signatures of major mergers.} However, the measured merger fractions at this epoch remain heavily dependent on the identification method employed, as different techniques are sensitive to different merger stages, mass ratios, and gas fractions \citep{lotz_major_2011, lackner_late-stage_2014}. \cite{hung_comparison_2014} demonstrated this directly by artificially redshifting local (U)LIRGS to $z \sim 1$ and found that $\sim$15\% lose their interacting signatures due to surface brightness dimming, while $\sim$30\% of (U)LIRGs at $z > 1$ are too faint in optical imaging for reliable classification altogether. As a result, merging systems that lack clear morphological signatures may be misclassified, suggesting that the contribution of mergers at $z \sim 1$ may be underestimated.

The launch of \textit{JWST} in 2021 marked a transformative era in our ability to study distant galaxies, offering unparalleled sensitivity and angular resolution in the infrared. With its 6.5 m primary mirror, \textit{JWST} achieves significantly greater resolving power compared to \textit{Spitzer} and \textit{HST}. Near Infrared Camera (NIRCam) observations have uncovered a greater number of disk galaxies at $z \sim 3-6$, approximately 10 times more than those identified in previous \textit{HST} observations \citep{ferreira_panic_2022}. Furthermore, the stellar mass and star formation rate (SFR) densities at $\nomStellMass > 10^{9} \nomSolMass$ are reported to be dominated by disk galaxies up to $z = 6$, suggesting that most stars in the universe likely formed in disk galaxies \citep{ferreira_jwst_2023}. Building on these findings, \cite{2025ApJ...994...94T}'s study of progenitors of Milky Way Analogs at $0.3 < z < 5$ from the Canadian NIRISS Unbiased Cluster Survey confirms that at higher redshifts, disk galaxies exhibit increased SFR and asymmetry. Moreover, the rate of double-peak mergers and disturbances to galaxy structure increases with redshift, classifying $\sim50\%$ of galaxies at $4 < z < 5$ as disturbed, and $\sim20\%$ as ongoing mergers. These findings demonstrate how the observing power of \textit{JWST} can uncover detailed structural features of galaxies that were previously unresolved, advancing our understanding of the mechanisms driving galaxy formation and evolution.

\begin{figure}[ht!]
\centering
\includegraphics[width=0.45\textwidth, height=\textheight, keepaspectratio]{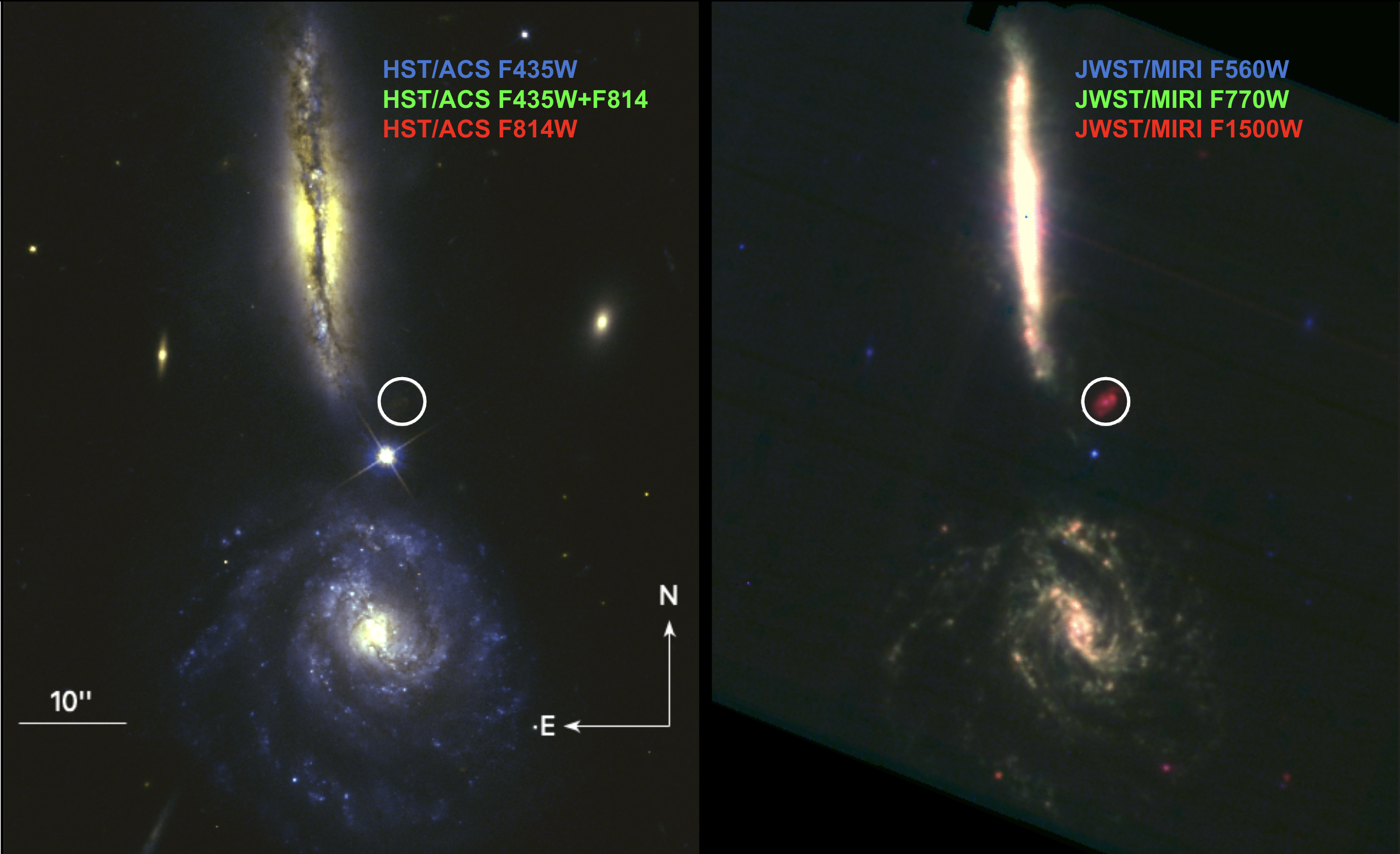}
\caption{\label{fig:hstjwst}(Left) \textit{HST} ACS RGB color image of VV 340 in F435W and F814W filters. (Right) \textit{JWST} MIRI RGB color image of VV 340 in F560W, F770W, and F1500W filters, with the galaxy of interest enclosed in the white circle. Faint emissions are barely visible in the \textit{HST} image, but the galaxy appears quite red in the MIRI image.}
\end{figure}

Given the unprecedented findings from \textit{JWST} imaging data of distant galaxies, which report higher merger rates than those previously observed, further examination of \textit{JWST} data is critical to validate these results and better understand the physical nature of these systems. This is particularly important because morphological classifications are often wavelength-dependent and subject to significant interpretative degeneracies between clumpy star-forming disks and genuine minor mergers \citep[e.g.,][]{guo_clumpy_2015, cibinel_early-_2019}. In this study, we examine an intermediate-redshift source (R.A. (J2000) 14:57:00.1086; Dec (J2000) +24:36:45.887; hereafter referred to as the ``lil gal'' as enclosed in the white circle in Figure \ref{fig:hstjwst}) serendipitously detected in MIRI imaging of the nearby system VV 340 as a method-demonstration for resolving these degeneracies. While the source is barely visible in \textit{HST} ACS imaging (F814W), it exhibits a complex, filter-dependent morphology in the mid-infrared (MIR) bands. \added{Specifically, the transition from an apparently isolated disk at F560W (5.6 $\mu$m) and F770W (7.7 $\mu$m) to a dual-peaked structure at F1500W (15 $\mu$m) presents a compelling case study for the necessity of multi-wavelength verification process.}

The goal of this paper is to determine whether such MIR morphological shifts represent a chance superposition of an interloper or a merger. By performing a case study into this single representative source through a combination of high-resolution GALFIT modeling, multi-band spectral energy distribution (SED) fitting, and Keck spectroscopy, we provide a robust framework for classifying the hundreds of thousands of similar star-forming galaxies being uncovered in current \textit{JWST} surveys. Our analysis highlights the nuances required to distinguish internal galactic processes from external triggers at the epoch of cosmic assembly.

This paper is structured as follows: Section \ref{data} details the \textit{JWST}, \textit{HST}, and Keck data used in our analysis. Section \ref{galfit} presents the GALFIT modeling of the observed components. Section \ref{photom} details the methods for photometry measurements and SED fitting. In Section \ref{keckmethods}, we describe the Keck spectral analysis. Finally, we present the interpretation and discussion of our results in Section \ref{results}. In our analysis, we adopt a cosmology of \(H_0 = 69.6\) km s\(^{-1}\) Mpc\(^{-1}\), \(\Omega_M = 0.286\), and \(\Omega_\Lambda\) = 0.714 \citep{wright_cosmology_2006}.

\section{Data}\label{data}
\subsection{James Webb Space Telescope Images}\label{jwstdata}
\begin{figure*}[ht]
\centering
\includegraphics[width=\textwidth, height=\textheight, keepaspectratio]{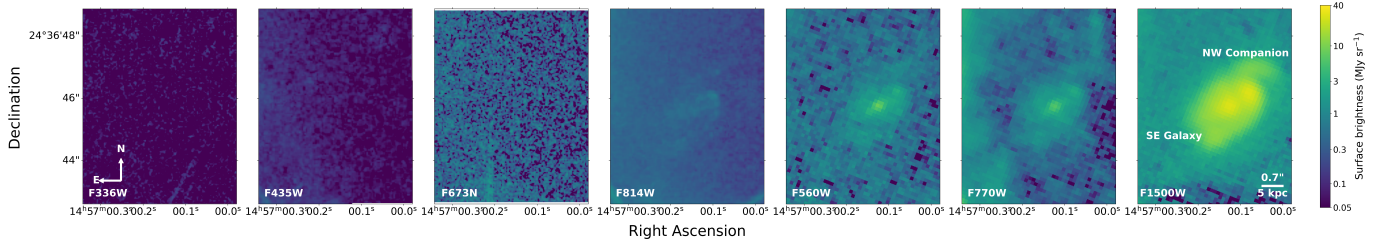}
\caption{\label{fig:miri}\textit{HST} images of the galaxy in F336W, F435W, F673N, and F814W filters; \textit{JWST} MIRI images of the galaxy in F560W, F770W, and F1500W filters. The source is barely discernible in the \textit{HST} filters, with faint traces of spiral arms observed in the F814W image. We observe a disk shape with a central bulge and spiral arms in F560W and F770W, with a faint second source. The galaxy appears brighter at increasing wavelengths, with a prominent second source in F1500W, NW of the central bulge. A constant background level (1st-percentile pixel value) was subtracted from each JWST/MIRI panel for display purposes.}
\end{figure*}

The MIRI imaging observations were made on 2023-03-24 UT as part of the GO-01717 program (PI: U), which focuses primarily on integral field unit (IFU) observations of nearby infrared galaxies in medium resolution spectroscopy (MRS) mode~\cite[][and other forthcoming papers]{kader_precessing_2026}. The MIRI imager was strategically placed on the target galaxy VV 340 while MRS pointed at nearby sky as background observations. As a result, the imaging data presented here were ``free images'' with relatively shallow exposure times (see Table~\ref{tab:exptime}) compared to those observations dedicated to high-redshift studies. Nonetheless, given the superb sensitivity of \textit{JWST} and with our chosen set of filters, we were able to identify ``lil gal'' from its unusually red appearance and intriguing multi-band morphology (Figure~\ref{fig:miri}), previously hidden in the diffuse background light of the foreground galaxy. We retrieved from the Mikulski Archive for Space Telescopes (MAST) the Stage 3 calibrated exposures processed by the \textit{JWST} Science Calibration Pipeline version 1.9.6 \citep{bushouse_jwst_2023}, in units of MJy sr$^{-1}$. 
\begin{deluxetable}{lccc}[hb!]
\tablecaption{\label{tab:exptime} \textit{HST} and \textit{JWST} Imaging Data Observations}
\tablehead{
    \colhead{Camera} & \colhead{Filter} & \colhead{Exposure Time (s)} & \colhead{Proposal ID}
}
\startdata
\textit{HST} WFC3/UVIS & F336W & 2349 & 16914 \\
\textit{HST} ACS/WFC & F435W & 1260 & 10592 \\
\textit{HST} WFC3/UVIS & F673N & 1800 & 14095 \\
\textit{HST} ACS/WFC & F814W & 720 & 10592 \\
\textit{JWST} MIRI & F560W & 1146 & 01717 \\
\textit{JWST} MIRI & F770W & 1146 & 01717 \\
\textit{JWST} MIRI & F1500W & 1146 & 01717 \\
\enddata
\end{deluxetable}

\subsection{Hubble Space Telescope Images}\label{hstdata}
Foreground galaxy VV 340 was previously observed by \textit{HST} with Wide Field Channel (WFC) of the Advanced Camera for Surveys (ACS) and Wide Fields Camera 3 (WFC3)/UVIS. 
The F435W \added{(0.43 $\mu$m)} and F814W \added{0.81 $\mu$m)} images were taken as part of PID 10592 (PI: Evans) on 2009-03-12 UT, while the F336W \added{(0.34 $\mu$m)} and F673N \added{(0.67 $\mu$m)} images were taken as part of PIDs 16914 (PI: Evans) and 14095 (PI: Brammer), respectively. The science-ready \textit{HST} images were downloaded directly from MAST.

\subsection{Keck Spectra}\label{keckdata}
Keck NIRES observations of ``lil gal'' were taken on 2023-05-08 (PI: Canalizo; PID: 2023A\_U179). The NIRES spectrograph covers 5 overlapping orders, 7 to 3, at an average spectral resolution of R \(\sim\) 3400 (\(\Delta\)$v$ \(\sim\) 84-89 km s\(^{-1}\)) with a 0.\arcsec55 $\times$ 18\arcsec slit \citep{wilson_mass_2004}. The NIRES wavelength coverage ranges from 0.94 to 2.45 \(\mu\)m, with a gap between 1.85$-$1.88 \(\mu\)m. Individual exposures for the target were taken for 4 minutes using the standard ABBA nodding, amounting to a total exposure time of 60 minutes. An A0 spectral class star was used as a telluric standard star, with measured magnitudes in J, H, and K bands. Observations of the telluric star were taken before and after the science target to correct for the atmospheric absorption features. Refer to \cite{aravindan_closer_2024} for details on the data reduction process. 

\begin{figure}
    \centering
    \includegraphics[width=\linewidth]{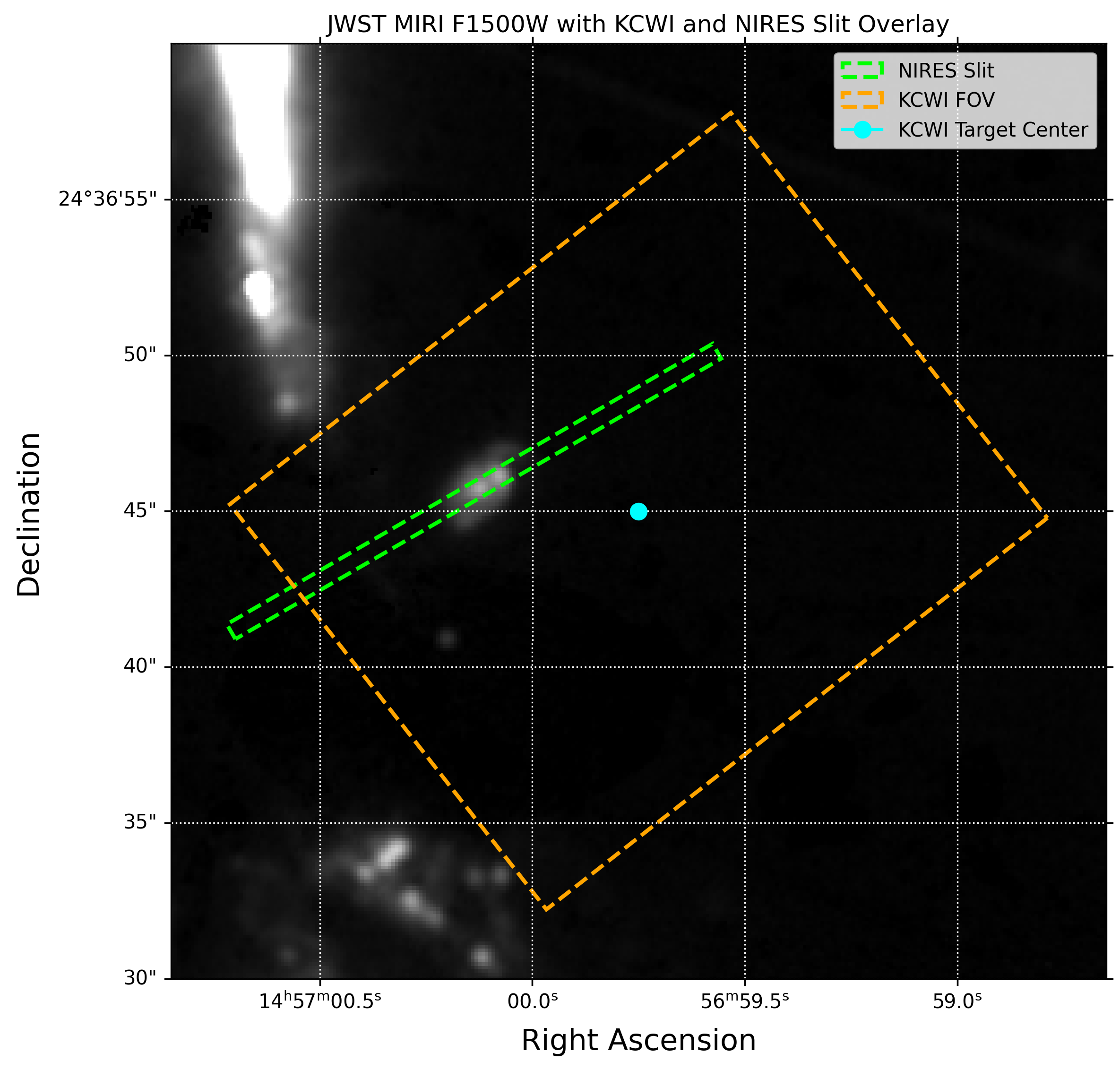}
    \caption{MIRI F1500W image overlaid with the Keck NIRES (green) slit position (slit center = R.A. (J2000) 14:57:00.1373; Dec (J2000) +24:36:45.633; PA = 30$\degree$) and KCWI (orange) FOV targeting ``lil gal''.}
    \label{fig:keck_fov}
\end{figure}

We supplement our data set with observations from Keck/KCWI, which provides optical integral field spectroscopy and captures key diagnostic lines beyond the wavelength range of the near-infrared slit spectra from NIRES. Follow-up KCWI and Keck Cosmic Reionization Mapper (KCRM) observations were taken on 2024-05-01 UT (PI: U; PID: 2024A\_U118). We produced three data cubes, each composed of three 300 s integrations. The medium slicer was used, which features a field of view (FOV) of 16\arcsec $\times$ 20\arcsec~(see Figure \ref{fig:keck_fov} for KCWI FOV and NIRES slit position) and a nominal spectral resolution of R \(\sim\) 2800 (\(\Delta\)$v$ \(\sim\) 100 km s\(^{-1}\)). With both the blue and red arms, KCWI+KCRM's wavelength coverage extends from 350 to 1050 nm. The red central wavelength was set to 8900 \text{\AA}, targeting the \hb~and \oiii~complex based on the redshift determination in Section \ref{nires_sec}. The data was reduced using the KCWI Data Reduction Pipeline (DRP)\footnote{\url{https://github.com/Keck-DataReductionPipelines/KCWI_DRP/releases}} and flux calibrated with the standard star Feige 34. 

\section{Morphological Analysis}\label{galfit}
We conduct a morphological analysis of ``lil gal'' to determine its structural parameters. The MIRI images in Figure \ref{fig:miri} reveal a disk galaxy with spiral arms and a central bulge at 5.6 and 7.7 \(\mu\)m. However, at 15 \(\mu\)m, a second peak emission appears NW of the central bulge, at an angular separation of 0.\arcsec7. We have labeled the second source as the NW Companion to the underlying SE galaxy. 

We apply GALFIT \citep{peng_detailed_2002} to all three bands of MIRI images and model the subcomponents in order to quantify their morphologies. We first crop the data images to yield a FOV of 7.\arcsec26 $\times$ 10.\arcsec01 
centered on ``lil gal'', ensuring sufficient background around it for a reliable background sky estimate. Upon observation, the galaxy exhibits a disk morphology with two distinct light sources. Therefore, a Sérsic profile enclosing the central disk and two Gaussian profiles centered around the central bulge and NW source are passed through GALFIT for all three MIRI images. The choice of Gaussian models for the NW and SE peaks was motivated by their compact nature. Inspection of the F1500W image with the QFitsView \footnote{\url{https://www.mpe.mpg.de/~ott/dpuser/qfitsview.html}} viewing tool shows that pixel slices through both peaks are well approximated by Gaussian profiles, which also reduce the number of free parameters compared to Sérsic models. We also provide a sky background parameter for each image. Since GALFIT is optimized to obtain estimates from ADUs of counts, we scaled the data images by the flux conversion factor from MJy sr$^{-1}$ to counts s$^{-1}$, obtained from the MIRI header keyword \(\mathrm{PHOTMJSR}\), and the exposure time. The AB magnitudes are then estimated by defining the zeropoint parameter (equation \ref{zp_ap}) from \cite{iani_first_2022}, where \(\mathrm{PIXAR\_SR}\) represents the nominal pixel area in steradians.
\begin{equation}\label{zp_ap}
\small ZP_{\text{AB}} = -2.5 \log_{10} (\mathrm{PIXAR\_SR} \cdot \mathrm{PHOTMJSR})-6.1
\end{equation}
All parameters were left free during the fitting, with the exception of the FWHM of the SE Gaussian component in the F560W image, which was held fixed at 2.5 pixels due to the faintness of the component at this wavelength. Galfit performs a Levenberg--Marquardt least-squares minimization to converge on the best-fit parameter values, using the reduced $\chi^2$ as the goodness-of-fit metric. The reduced $\chi^2$ values returned by GALFIT for each MIRI band are reported and discussed in \S\ref{galf_results}.

\begin{deluxetable}{lccc}[hb]
\tablecaption{\label{tab:galfit_params} GALFIT Parameter Results of \textit{JWST} MIRI Data}
\tablehead{
    \colhead{Parameter} & \colhead{F560W} & \colhead{F770W} & \colhead{F1500W}
}
\startdata
Sérsic Index ($n$) & 1.22 \(\pm\) 0.0966 & 0.239 \(\pm\) 0.0411 & 0.472 \(\pm\) 0.0166 \\
SE Mag & 21.4 \(\pm\) 0.159 & 20.8 \(\pm\) 0.115 & 19.5 \(\pm\) 0.229 \\
NW Mag & 22.4 \(\pm\) 0.137 & 21.9 \(\pm\) 0.311 & 18.9 \(\pm\) 0.0495 \\
\enddata
\end{deluxetable}

\begin{figure}[h!]
\centering
\includegraphics[width=0.5\textwidth, height=\textheight, keepaspectratio]{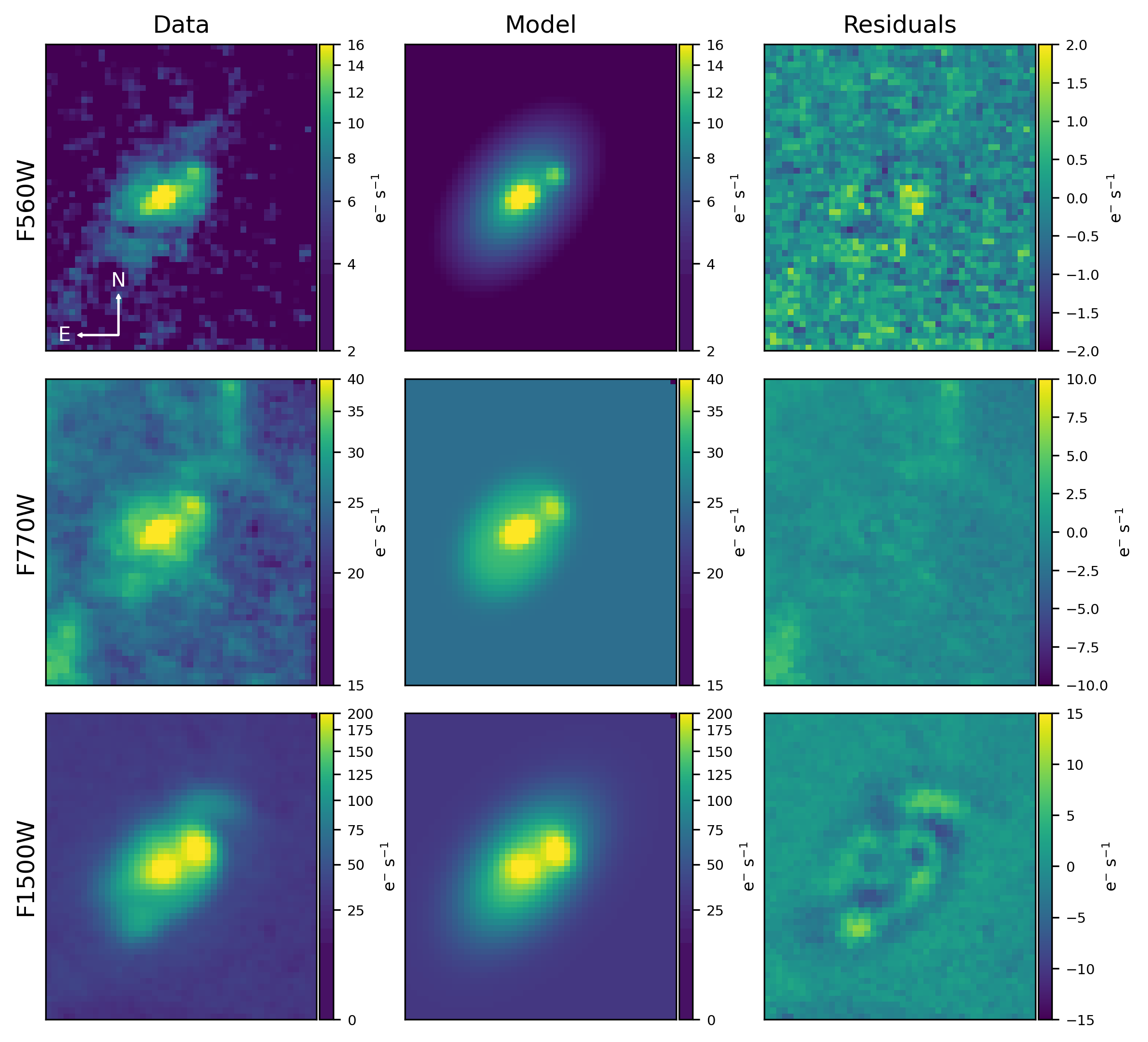}
\caption{\label{fig:galf}GALFIT results of the MIRI image filters in F560W, F770W, and F1500W (top to bottom). Each row includes the data input image, the modeled components, and the residuals. Each model includes a Sérsic profile enclosing the central disk and two Gaussian profiles for the central bulge and NW companion.}
\end{figure}

 We present GALFIT parameter estimates of the Sérsic and Gaussian profiles in Table \ref{tab:galfit_params}. \added{We report a Sérsic index of $n \sim 1.22$ in the F560W image, consistent with a galactic disk, and lower values in F770W and F1500W associated with irregular morphologies.} The GALFIT modeling results are shown in Figure \ref{fig:galf}, including the input image (left), the model (middle), and the residuals (right) for all three MIRI filters and are discussed further in \S\ref{galf_results}.

\section{Photometry Measurements}\label{photom}
To determine the appropriate aperture size for measuring the photometry of ``lil gal,'' we followed a curve of growth method using PHOTUTILS \citep{bradley_astropyphotutils_2023} on the MIRI F1500W image, where the features are most prominent. We also considered background contamination in the F770W image in the aperture size selection, since we observed prominent 7.7 $\mu$m polycyclic aromatic hydrogen (PAH) emissions from VV 340 (see \S\ref{galf_results} for details). 
To balance capturing the source flux while minimizing background contribution, we adopted a circular aperture with a radius of 2.\arcsec163. This size reflects the point at which the curve of growth of the source flattens in the F1500W image, while avoiding regions of prominent contamination from VV 340 in the F770W image. The same aperture was applied consistently across all imaging bands.

\subsection{\textit{HST} Photometry}\label{hstphotmethods}
The flux measurements are obtained using the PHOTUTILS \texttt{SkyCircularAperture} routine. To account for the background sky, we measured the background level from several regions along the isophotes surrounding VV 340 at the distance of ``lil gal'' in order to best estimate the foreground contamination level at the source of interest. The flux is converted to mJy by multiplying the sky-subtracted flux in units of electrons s$^{-1}$ by the inverse sensitivity at the infinite aperture indicated by the keyword PHOTFLAM in units of ergs cm$^{-2}$ \text{\AA}$^{-1}$ electron$^{-1}$, the square of the pivot wavelength in Angstroms (PHOTPLAM), and dividing by the encircled energy ($EE$) fraction. Here, $EE=1$ due to the large aperture size. Flux measurements are reported in Table~\ref{tab:photflux}, where we observe increasing fluxes at increasing wavelengths.
\begin{deluxetable}{lccc}
\tablewidth{0pt}
\tablecaption{\textit{HST} and \textit{JWST} Photometry\label{tab:photflux}}
\tablehead{
  \colhead{Filter} & \colhead{Total System\tablenotemark{a}} & \colhead{NW} & \colhead{SE} \\
  \colhead{} & \colhead{Flux (mJy)} & \colhead{Flux (mJy)} & \colhead{Flux (mJy)}
}
\setlength{\tabcolsep}{3pt}
\decimals
\startdata
F336W & 0.00135 $\pm$ 0.000941 & \nodata & \nodata \\ 
F435W & 0.00649 $\pm$ 0.00484 & \nodata & \nodata \\
F673N & 0.0263 $\pm$ 0.0167 & \nodata & \nodata \\
F814W & 0.0318 $\pm$ 0.0165 & \nodata & \nodata \\
F560W & 0.731 $\pm$ 0.330 & 0.00411 $\pm$ 0.000727 & 0.00977 $\pm$ 0.00189 \\
F770W & 1.93 $\pm$ 1.19 & 0.00573 $\pm$ 0.00131 & 0.0167 $\pm$ 0.00375 \\
F1500W & 17.2 $\pm$ 4.50 & 0.0949 $\pm$ 0.0244 & 0.0562 $\pm$ 0.0146 \\
\enddata
\tablenotetext{a}{The Total System encloses the NW and SE components, along with the spiral arms that were not included in the GALFIT model.}
\end{deluxetable}

\subsection{\textit{JWST} Photometry}\label{jwstphotmethods}
The Space Telescope Science Institute (STScI) provides notebooks for post-pipeline analysis of \textit{JWST} data, including aperture photometry for MIRI Stage 2 images.\footnote{\url{https://www.stsci.edu/jwst/science-execution/jwebbinars}} We make use of the steps detailed in these notebooks as a guide for our analysis. 
Since we retrieved Stage 3 pipeline-calibrated data, we do not apply the additional calibration steps provided by the script.  

We begin with a background subtraction utilizing the \texttt{Background2D} function in the PHOTUTILS package. This method constructs a 2D background model from NxM subregions, from which the background level in each region is estimated to produce a background image with a median filter. For source identification, the STScI notebook uses \texttt{DAOStarFinder}, which locates sources by identifying local density maxima. Since our study focuses on a single galaxy, we pass the galaxy’s xy pixel coordinates directly to \texttt{DAOStarFinder} to bypass the automatic detection of multiple sources and focus our analysis on the object of interest. We then apply the PHOTUTILS aperture photometry routine, similar to the process described in \S\ref{hstphotmethods}. We define an aperture radius of 19.6 MIRIM pixels, which we determined by applying the MIRI conversion factor of 0.\arcsec11 px$^{-1}$ to the aperture radius of 2.\arcsec163 used in \S\ref{hstphotmethods}. To account for any remaining background from the \texttt{Background2D} model subtraction, we subtract the sum within a circular annulus of inner and outer radii of 25 and 35 pixels from the sum within the aperture surrounding the source.

\begin{figure}[h!]
\centering
\includegraphics[width=0.5\textwidth, height=\textheight, keepaspectratio]{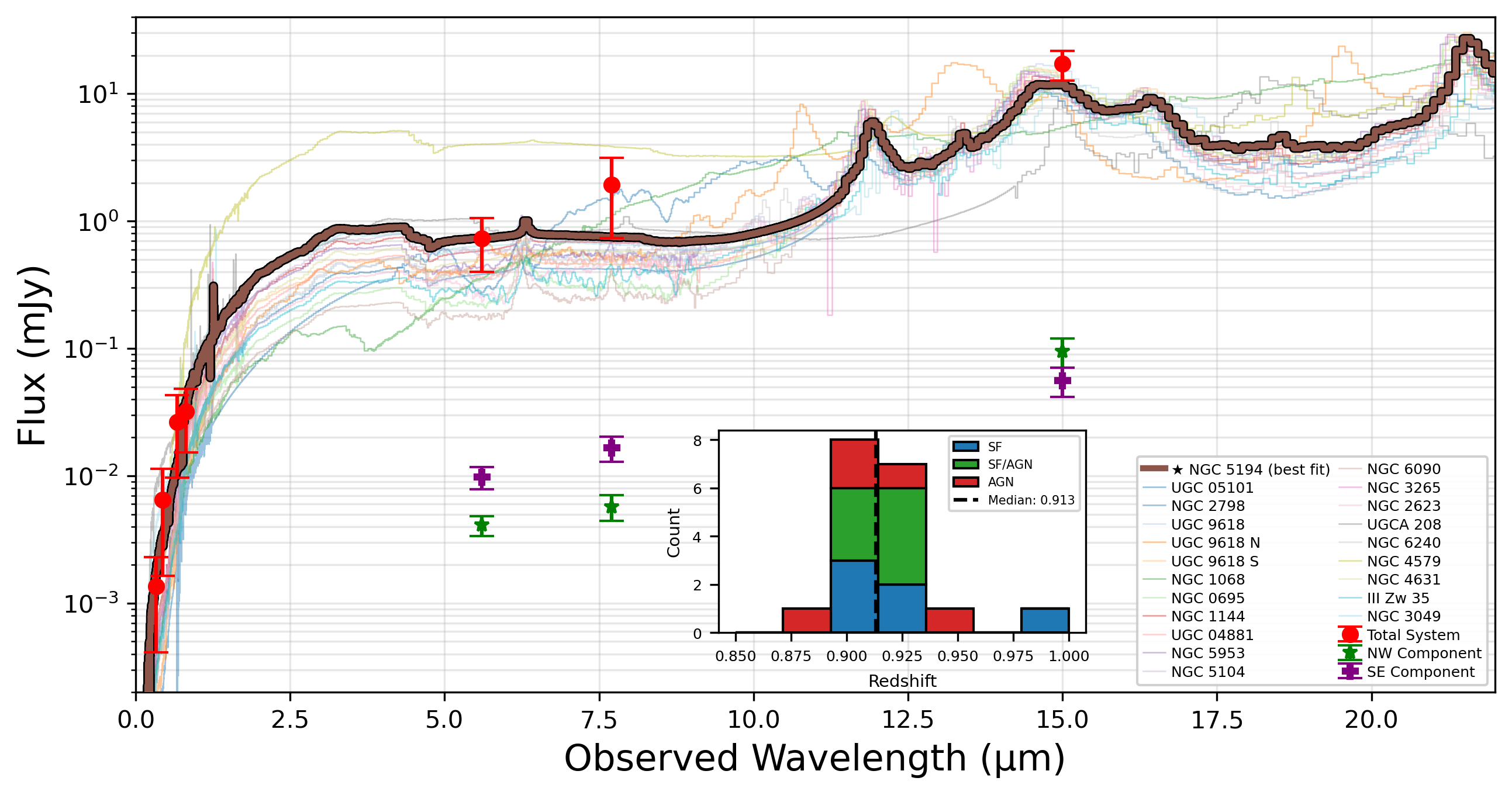}
\caption{\label{fig:photom}Photometry results from the total system for all \textit{HST} and \textit{JWST} images (red), fitted with SED templates from \cite{brown_atlas_2014}, with NGC 5194 (bold) shown as the best-matching template. \textbf{While the median redshift across all the templates is 0.913,} the photometry is best matched by SF and SF/AGN composite templates at $z$ \(\sim\) 0.92. Decomposed fluxes from the NW component (green) and SE component (purple) from the three MIRI images follow the same trend of the total system at increasing wavelengths.
}
\end{figure}

The MIRI pixel unit for Stage 3 images is MJy sr$^{-1}$. To convert the background-subtracted photometry calculations to units of mJy, we multiply by the average pixel area in steradian from the header keyword \(\text{PIXAR\_SR}\) and additional scaling conversions. The resulting fluxes are reported as the Total System Flux in Table \ref{tab:photflux}, along with the NW and SE components from the GALFIT-generated subcomponent models. All \textit{HST} and \textit{JWST} flux measurements are plotted in Figure \ref{fig:photom}, and are discussed in further detail in \S\ref{sed_results}.

\subsection{Spectral Energy Distribution Fitting}\label{sedmethods}
We perform \added{Spectral Energy Distribution (SED)} fitting to estimate the photometric redshift of ``lil gal'' and to explore the physical nature of the system by comparing its observed emission to templates of star-forming (SF) and AGN-hosting galaxies. The \textsc{GALSEDATLAS} \citep{brown_atlas_2014} provides a catalog of 129 UV--MIR SEDs for a variety of galaxy populations. We tested templates classified as SF, AGN-hosting, and composites, further narrowing our selection to 21 templates with morphological characteristics similar to ``lil gal,'' including peculiar and spiral types. For each template, we normalized the SED to match the integrated observed flux over the mid-infrared wavelength range, where the MIRI photometry has the highest signal-to-noise, and varied to find the best-fit $\chi^2$ computed in log-space, since the observed fluxes span approximately four orders of magnitude. The reduced $\chi^2$ measures how well each redshifted template matches the observed photometric points. The best-matching template, NGC 5194, yields $z = 0.92$ with reduced $\chi^2 = 1.16$, close to unity, indicating a good fit within the photometric uncertainties. The five templates with acceptable fits ($\chi^2 < 2$) all converge to $z = 0.91$--$0.98$, and are consistently SF and SF/AGN composites. \added{The median redshift across all 21 templates is 0.913. We adopt $z \sim 0.92$ as our photometric redshift, based on the best-fit template NGC 5194.} Figure~\ref{fig:photom} shows all 21 templates at their best-fit redshifts.

\section{Spectroscopy Measurements}\label{keckmethods}
\subsection{Keck NIRES Spectra}\label{nires_sec}
\begin{figure*}[ht]
\centering
\includegraphics[width=\textwidth, height=\textheight, keepaspectratio]{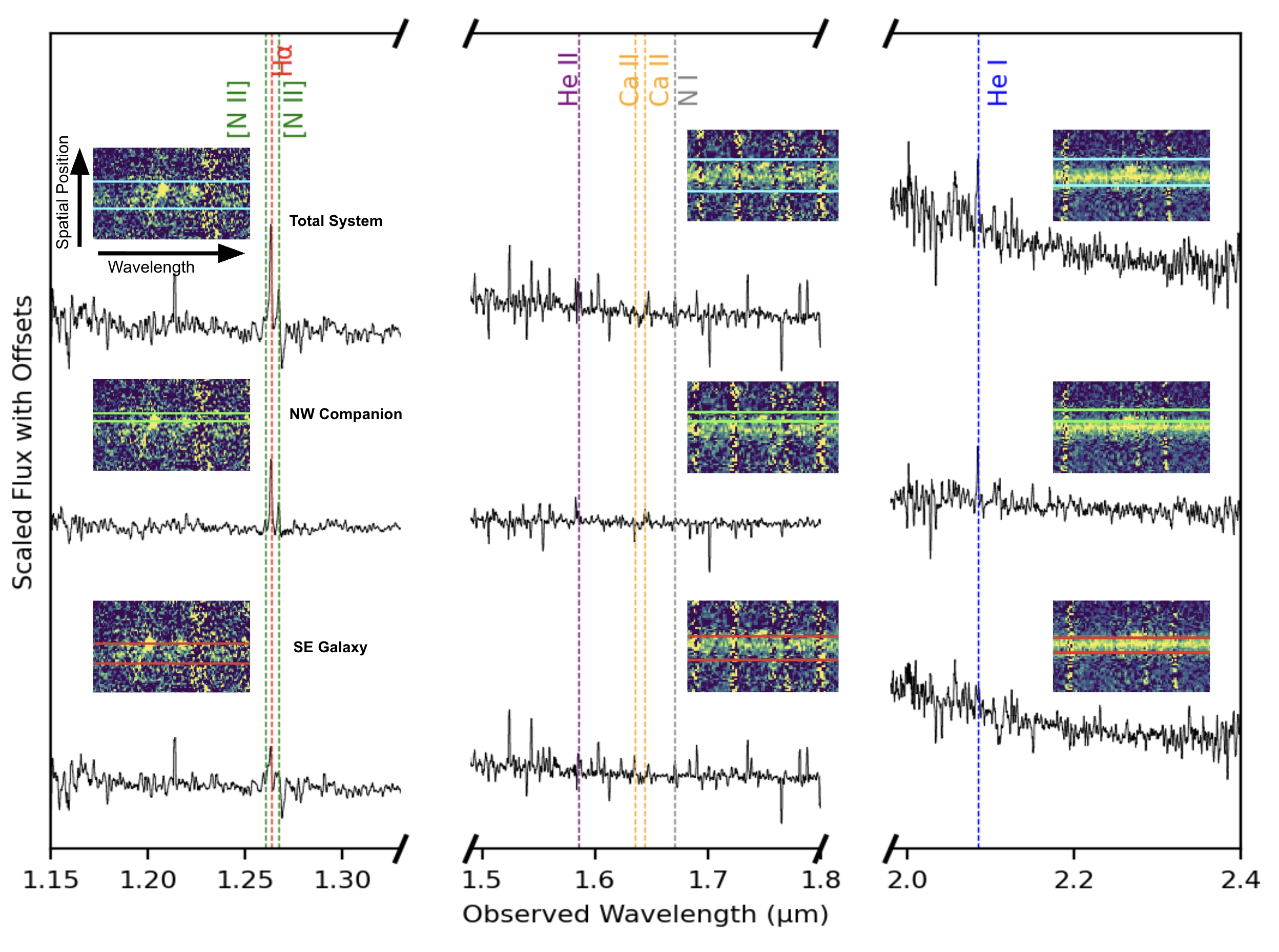}
\caption{\label{fig:spectra_ext_thesis}Keck NIRES spectra plot of the system from order 5 (left), order 4 (middle), and order 3 (right). Each order's extraction is split according to the total system (top row), the NW companion (middle row), and the SE galaxy (bottom row). Horizontal lines overlaid on images of the 2D spectra on the left represent the selected regions for extraction. Emission lines corresponding to $z=0.9248$ are detected, primarily from the NW companion source. For presentation purposes, we smoothed spectra using a kernel width of 6. 
}
\end{figure*}

Of the five available orders of NIRES data, we focus on orders 5 through 3 (1.1-2.45 \(\mu\)m) since orders 7 and 6 are featureless. The NW and SE components are spatially resolved in the 2D NIRES data, with the SE component appearing as a horizontal continuum trace and the NW component as a compact emission feature offset along the slit direction. The 2D spectral insets in Figure~\ref{fig:spectra_ext_thesis} show these features with horizontal lines marking the extraction regions for the total system, NW component, and SE component. We perform the subsequent spectral analysis by separately extracting these three apertures, as shown in Figure~\ref{fig:spectra_ext_thesis}. The $\sim$0.7$''$ angular separation between the two sources corresponds to $\sim$5 NIRES pixels at the spatial pixel scale of 0.\arcsec15 per pixel. Note that the offsets between the orders may be attributed to uncertainties in the flux calibration and background subtraction.

\begin{figure*}[ht!]
\centering
\includegraphics[width=0.7\textwidth, height=\textheight, keepaspectratio]{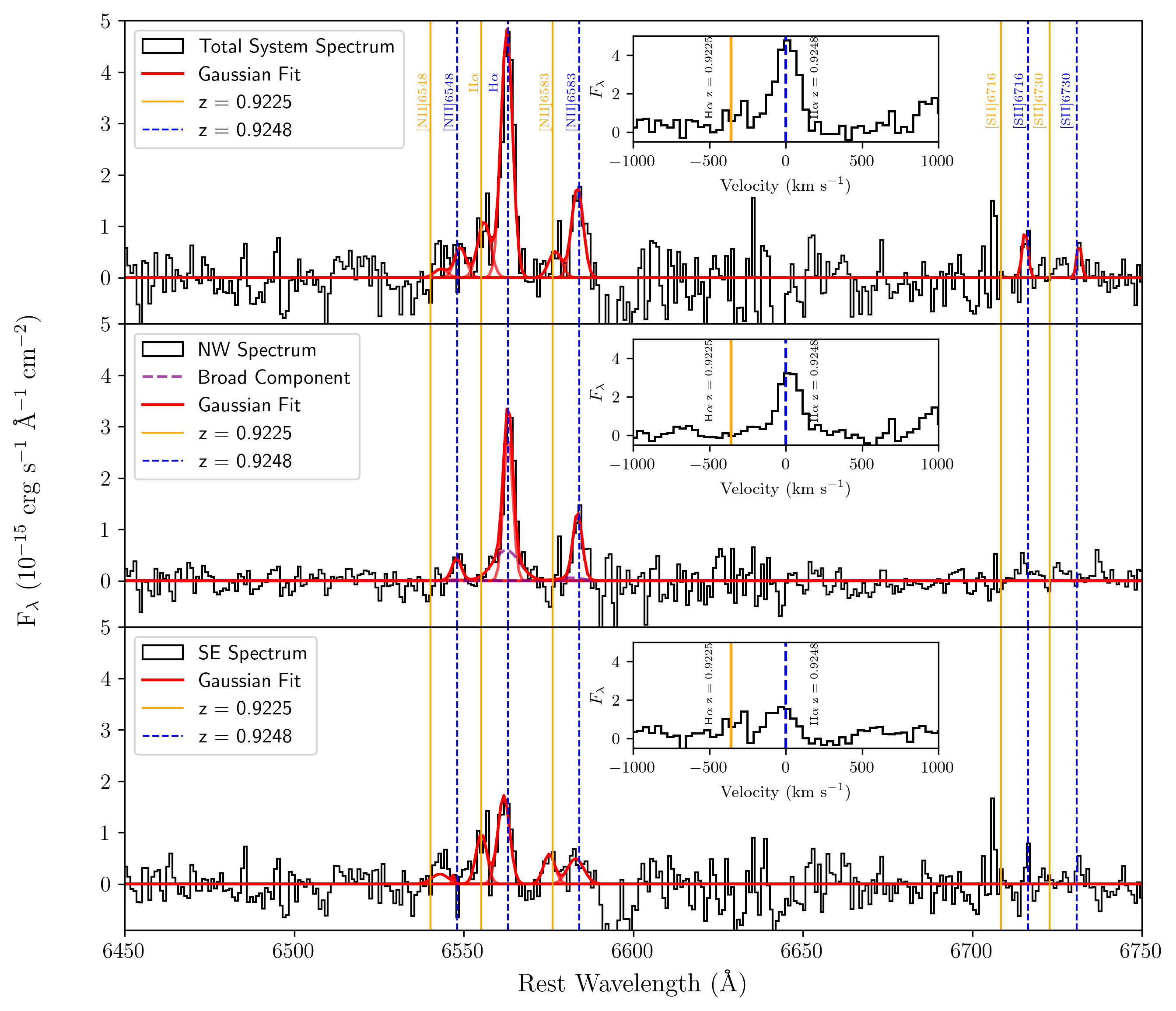}
\caption{\label{fig:linefit_combined}Order 5 NIRES spectrum extracted from the total system (top), NW component (middle), and SE component (bottom). The total extraction and SE extraction is fitted with a six-component compound Gaussian model for the two sets of distinct peaks corresponding to \nii~6548 \text{\AA}, \ha, and \nii~6583 \text{\AA} at z = 0.9225 (orange) and z = 0.9248 (blue). The NW extraction is fitted with a six-component Gaussian extraction, comprised of broad and narrow components for \nii~6548 \text{\AA}, \ha, and \nii~6583 \text{\AA} at z = 0.9248. The inset in the total and SE extraction highlights the \ha~velocity separation of 360 km s\(^{-1}\) between the NW and SE components.} 
\end{figure*}

We observe the strongest emission lines in the order 5 spectrum, and identify \nii~6548 \text{\AA}, \ha, and \nii~6583 \text{\AA} at $z$ = 0.9248 $\pm 0.0006$, from the NW extraction. Blue-shifted emission lines corresponding to $z$ = 0.9225 $\pm 0.0006$ are observed in the total system's spectrum, highlighted in Figure \ref{fig:linefit_combined}. 

To calculate emission line flux ratios for subsequent ionization diagnostics in \S\ref{bpt}, we performed Gaussian 
fitting to the detected emission lines in the NIRES and KCWI spectra. 
Prior to fitting, a linear baseline was subtracted from each spectrum, 
estimated from nearby continuum regions on either side of the 
emission features. We then fit compound Gaussian models to the 
continuum-subtracted spectra using Astropy's Trust Region Reflective 
least-squares fitter (\texttt{TRFLSQFitter}). Integrated line fluxes, reported in Table \ref{tab:linefit_results}, were computed from the Gaussian profiles using the \texttt{line\_flux} routine in the SPECUTILS package \citep{earl_astropyspecutils_2024}.

For the NIRES order 5 spectrum, we fit the \nii~and \ha~emission lines using compound Gaussian models. In the total system and SE extractions (top and bottom rows of Figure \ref{fig:linefit_combined}), each line at $z = 0.9225$ and $z = 0.9248$ is fit with a single Gaussian. In the NW extraction (middle row), the lines at $z = 0.9248$ are fit with both a broad and narrow component, a choice motivated by the extended wings observed in \ha. The widths of the broad components were tied across the \nii~and \ha~lines, as were the narrow components. The \nii~doublet amplitudes were fixed according to the theoretical 1:2.96 ratio. We report a width of $\sigma \sim 177.72 \pm 23.61 $km s$^{-1}$ for the \ha~broad component. Similarly, the \oiii~4959 \text{\AA} amplitude was fixed to the theoretical ratio of 1:2.98 relative to \oiii~5007 \text{\AA} \citep{storey_theoretical_2000}, and the \oiii~line widths were tied together.

Using the flux ratios derived from the NIRES and KCWI spectra, we then estimated gas-phase oxygen abundances ($12 + \log(\mathrm{O}/\mathrm{H})$) following multiple strong-line diagnostics \citep{pettini_o_2004, charlot_nebular_2001, denicolo_new_2002}. The metallicity methods and results are listed in Table~\ref{tab:metallicity} and are discussed further in \S\ref{metallicity}.

\subsection{Keck KCWI Spectra} \label{sec:style}
\begin{figure*}[ht!]
\centering
\includegraphics[width=\textwidth, height=\textheight, keepaspectratio]{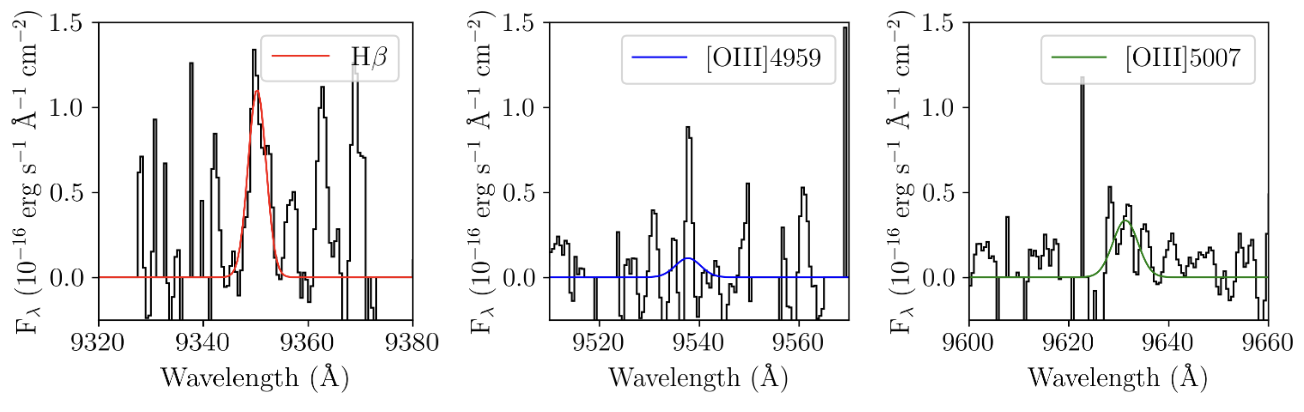}
\caption{\label{fig:kcwispectra}KCWI spectra of the total system, highlighting the \hb, \oiii~4959 \text{\AA}, and \oiii~5007 \text{\AA} emission lines, corresponding to $z \sim 0.9235$. For presentation purposes, we applied sigma-clipping of $\sigma = 2.5$ at 5 iterations to clip the prominent skylines surrounding the emission lines.
}
\end{figure*}

We supplement the NIRES data with observations from Keck's KCWI instrument to aid in classifying the observed sources. 
To optimize the signal of the KCWI spectra, we co-add the three cubes from the observing run, accounting for dithering and flux calibration. The process begins with three flux-calibrated and telluric-corrected data cubes. We spatially align the cubes at the subpixel level to correct for the dither position offsets determined from Astropy's \texttt{image\(\textunderscore\)registration} routine. 
The three spatially aligned cubes are then co-added. An extraction aperture of radius 2.\arcsec5 is used, based on our NIRES extraction size, and corrected for KCWI's rectangular pixel dimensions. The extracted spectrum of the total system is plotted in Figure \ref{fig:kcwispectra}, highlighting the \hb,~\oiii~4959 \text{\AA}, and \oiii~5007 \text{\AA} emission lines detected at an intermediate redshift of $\sim 0.9235$.
  
\section{Results and Discussion}\label{results}

\subsection{Galaxy Morphology and Structure}\label{galf_results}
From the GALFIT-generated parameter estimates in Table \ref{tab:galfit_params}, we observe changes in the structure of the galaxy with increasing wavelengths. The Sérsic index $n$ \(\sim\) 1 in the F560W image indicates a galactic disk, while the lower $n$ values from F770W and F1500W are associated with irregular and peculiar morphologies. The lower $n$ from the F770W model may have resulted from the significant background contamination observed from VV 340 since the F770W filter samples the 7.7~$\mu$m~PAH emission at rest wavelength for nearby galaxies. Similarly, the low $n$ from the F1500W model is likely attributed to the prominent NW companion source. Other studies have classified galaxies with $n$ \(\leq2\) as “pseudo-bulges” with enhanced star formation, tightly linked to nuclear starbursts triggered by mergers \citep{kormendy_secular_2004, he_close_2022}. The prominent emissions from both components in the 15 $\mu$m image --- particularly from the NW companion --- offers further insight into the nature of the source. At this wavelength, we are probing a rest-frame wavelength of $\sim$ 7.7~$\mu$m, where strong PAH emission is expected, a well-established signature of starburst activity in the mid-infrared \citep{genzel_what_1998, lutz_nature_1998}.

We report reduced \(\chi^2\) values of \(\sim\) 0.83, 1.83, and 2.97 for the fits corresponding to the F560W, F770W, and F1500W images, respectively. The relatively large $\chi^{2}$ value for the F1500W model likely arises from not fitting the spiral arms in the image. Additionally, the poorer fit in F1500W may support the merger hypothesis between the SE and NW components, rather than a single source with two components. This wavelength-dependent appearance has direct implications for merger identification in \textit{JWST} surveys. The NW companion brightens by $\sim$3.5 mag from F560W to F1500W, and while faintly visible at 5.6 and 7.7 $\mu$m, its appearance at these wavelengths is consistent with a star-forming clump within the disk rather than a distinct merging companion. Only at 15 $\mu$m does it emerge as the dominant source, revealing the dual-peaked structure that motivated this study. This underscores the importance of multi-wavelength coverage for achieving a complete census of close-pair mergers at $z \sim 1$. 

Similar studies examining the morphological properties of galaxies in the CANDELS fields can provide more insights into the nature of the source. A study by \cite{margalef-bentabol_formation_2016} using a sample of 1495 cosmic-noon (1 $< z <$ 3) galaxies finds that the number of two-component disk galaxies dominates, modeled by a Sérsic and exponential profiles. This study includes merger-like features in their single and double-component Sérsic models. The \cite{guo_clumpy_2015} study from CANDELS reports that minor mergers are a viable explanation for off-center clumps at \(z < 1.5\). \cite{zanella_contribution_2019} addressed the clump versus satellite classification for a sample of 53 $z \sim 1-3$ galaxies using a similar GALFIT decomposition approach and multi-property photometric criteria, but noted that spectroscopic confirmation of individual components was not available for most of their sample. While our GALFIT results offer tantalizing hints that ``lil gal'' may be a merger system in this context, morphological analysis alone cannot definitively rule out the clumpy disk or chance superposition scenarios --- the photometric and spectroscopic diagnostics presented in the following sections provide the additional evidence needed to break this degeneracy.

\subsection{Photometric Redshift}\label{sed_results}
From the SED fitting presented in Figure \ref{fig:photom}, we estimate a photometric redshift of $z \sim 0.92 \pm 0.03$, confirming that ``lil gal'' is a distant source. The best-fit template is NGC 5194 (M51), an SF/AGN composite interacting galaxy, with a reduced $\chi^2$ of 1.16 at $z = 0.92$. The remaining top-ranked templates are also SF and SF/AGN composites with reduced $\chi^2$ values of 1.2--1.5. The redshift distribution inset in Figure \ref{fig:photom} favors starburst and SF/AGN composite activity at this redshift. At $z \sim 1$, \cite{kaviraj_galaxy_2015} reports that approximately 30\% of massive galaxies have undergone a major merger, with minor mergers being even more frequent. \added{Additionally, the infrared hump at the $\sim 8 \mu$m rest wavelength is characteristic of dusty star-forming galaxies \citep{sanders_ultraluminous_1988, 2011A&A...533A.119E, u_spectral_2012}.}

We extend our analysis of the photometric properties of the NW component by comparing its infrared spectral index to those of starburst and AGN sources. From photometric measurements of the GALFIT MIRI models, we derive a spectral index of $\alpha_{2.9}^{7.8}$ = 0.93 \(\pm\) 0.13, using the equation reported in \cite{weedman_active_2006}: \begin{equation}
    \alpha = log(\frac{\nu_1 f_(\nu_1)}{\nu_2 f_(\nu_2)}).
\end{equation} Comparing the spectral index results to recent \textit{JWST} NIRSpec and MIRI observations of Cygnus A \citep{2025ApJ...983...98O}, a local radio galaxy with a supermassive black hole at the center, we calculate a spectral index for the nucleus of $\alpha_{2.9}^{7.8}$ = 1.06 $\pm$ 0.15. The authors report that the continuum emissions at $> 2 \mu$m are dominated by hot dust emissions from the AGN torus. The spectral slope estimates between the two sources are in agreement, but considering that Cygnus A is a local source ($z = 0.0577$), we extend this comparison to higher redshift sources more relevant to ``lil gal''. \cite{weedman_active_2006} report spectral index values derived from \textit{Spitzer} Infrared Spectrograph (IRS) observations of eleven AGNs \(\ (0.6 < z < 2.5)\) and nine starbursts \(\ (1.0 < z < 1.9)\). Note that we limit our comparison to nine AGNs due to evidence of a stellar component reported in two sources from the AGN sample. The spectral index values range between $\alpha_{2.3}^{8.3}$ = 0.07 $-$ 0.62 for AGNs and a narrower range of $\alpha_{2.8}^{8.5}$ = 0.56 $-$ 1.05 for starbursts. Based on these results, we find that the NW companion's higher slope value is consistent with infrared starburst populations. While photometric estimates provide an efficient method for analyzing the cosmological distance of extragalactic sources, 
we cannot deliver precise measurements or definitively classify components, necessitating detailed spectroscopic analyses to accurately characterize the spectral properties and fully determine the nature of the system.

\subsection{Spectroscopic Analysis}\label{spec_redshift}
Spectroscopic follow-up is essential for distinguishing among the three competing scenarios for ``lil gal,'' since imaging and photometry alone cannot break the degeneracy between the merger, clumpy-disk, and interloper interpretations at $z \sim 1$. The Keck NIRES and KCWI observations described below provide the redshift, kinematic, and ionization information needed to break this degeneracy and characterize the two components.

The NIRES spectra provide crucial insights into the system. Emission
lines extracted from the SE galaxy are considerably weaker compared to its NW counterpart, as seen in Figure \ref{fig:spectra_ext_thesis}. From the total extraction, the \ha~and \nii~$\lambda\lambda$ 6548,6583 features in the NW source are detected at 1.5, 12.7, and 4.6 -$\sigma$, respectively,
while in the SE source they are only detected at 0.5, 2.8, and 1.4-$\sigma$. While the \nii~6548 \text{\AA} line detection from the SE component is marginal, the \ha~and \nii~6583 \text{\AA} lines appear at a different
redshift ($z = 0.9225$) from the NW companion (at
$z = 0.9248$), highlighted in the total (top) and SE (bottom) extraction in Figure \ref{fig:linefit_combined}. At this redshift, the angular separation of 0.\arcsec7 is roughly 5 kpc in the projected separation between the NW and SE components, placing ``lil gal'' within the close-pair regime defined \citep{yuan_role_2010}, where merger-induced star formation enhancement is expected to peak. The difference between the central peaks of the two well-detected \ha~lines suggests a velocity offset of 360 km s$^{-1}$, having defined the systemic velocity at the position of the NW \ha~line. This separation is consistent across multiple detected emission lines including \nii~6583 \text{\AA} and He I 10830 \text{\AA}. This velocity difference lies within the criteria of \(\Delta\)$v$ \(\leq\) 500 km s\(^{-1}\) for gravitationally bound close pairs \citep{patton_new_2000, de_propris_millennium_2007}, lending strong kinematic evidence to the scenario where the NW and SE sources are two interacting components within a galaxy merger system. We further observed that the \ha~and \nii~6583 \text{\AA} lines associated with the NW component are also detected in the SE extraction, appearing blue-shifted by $\sim$ 60 km s$^{-1}$ in \ha,~
consistent with disk rotation. This provides additional support that the lines detected at $z = 0.9225$ originate from a separate source, as they appear kinematically distinct from the rotation.

\begin{figure*}
\centering
\includegraphics[width=\textwidth]{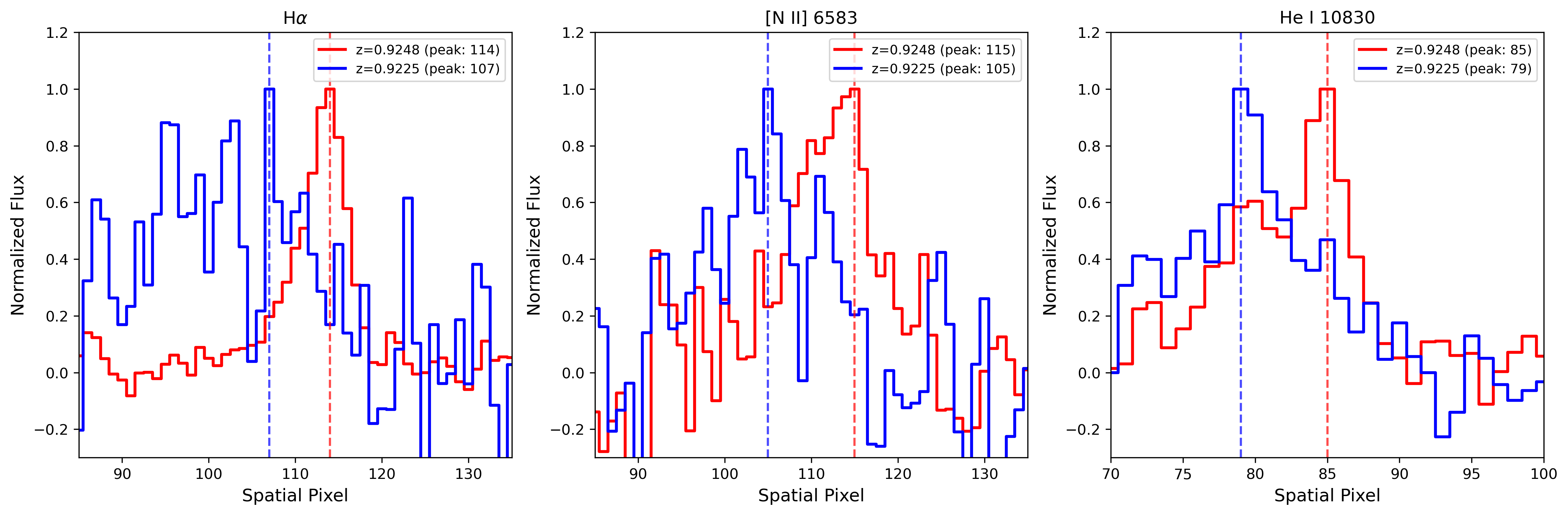}
\caption{Spatial profiles along the NIRES slit at the redshifted wavelengths of \ha~(left), \nii~6583 \text{\AA} (middle), and He I 10830 \text{\AA} (right). For each line, we integrate the flux in a narrow wavelength window centered on the line position at $z = 0.9248$ (red) and $z = 0.9225$ (blue), then sum along the wavelength axis to obtain flux as a function of spatial row along the slit. Dashed vertical lines mark the peak of each profile. The peaks at the two redshifts are offset by 6--10 pixels, confirming that the two velocity components are spatially distinct.}

\label{fig:spatial_profiles}
\end{figure*}

To verify that the NW and SE components are spatially distinct along the slit and not a single blended source, we extracted spatial profiles of the three emission lines. For \ha, \nii~6583, and He I 10830 \text{\AA}, we integrated the flux in a narrow wavelength window ($\pm$3 pixels for order 5, $\pm$1 pixel for order 3) centered on the line at each candidate redshift ($z = 0.9248$ and $z = 0.9225$), producing a spatial profile of line flux along the slit (Figure \ref{fig:spatial_profiles}). In all three lines, the emission at $z = 0.9248$ peaks at a spatial position offset from the emission at $z = 0.9225$ by 6--10 pixels. Although the $\sim$ 0.\arcsec5–0.\arcsec6 seeing during the NIRES observations is comparable to the $\sim$0.\arcsec7 projected separation --- producing some unavoidable PSF overlap in the wings --- the consistent offset in peak positions confirms that the two velocity components correspond to spatially distinct regions along the slit.

We can determine the SFR of the NW and SE components from their respective \ha~fluxes reported in Table \ref{tab:linefit_results}, converted to luminosities from their measured redshifts. The estimated log(SFR) for the NW and SE components, calculated from \cite{calzetti_star_2007}, are 2.37 $\pm$ 0.04 \nomSolMass~yr$^{-1}$ and 2.20 $\pm$ 0.09 \nomSolMass~yr$^{-1}$, respectively. \added{Compared to other studies of galaxies at similar epochs \cite[$0.5 < z < 2.5$;][]{noeske_star_2007, whitaker_constraining_2014, 2015A&A...575A..74S},} these SFRs exceed typical values by a factor of $\sim 10$. This excess may indicate ongoing merging activity, consistent with findings that major mergers can enhance star formation by factors of 10$–$100 relative to isolated galaxies of similar mass \citep{matteo_star_2007, matteo_frequency_2008}.

As an additional check, we compare the stitched spectrum from the SE extraction to stellar populations of varying ages to determine whether the system could be classified as a clumpy disk. \cite{guo_clumpy_2015} studied a sample of star-forming galaxies from CANDELS and reported peak fractions of clumpy galaxies at cosmic noon $1 < z < 3$. Previous \textit{HST} and more recent \textit{JWST} studies report younger stellar age ranges of clumps from tens of Myr to hundreds of Myr, reaching up to 1 Gyr \citep{guo_multi-wavelength_2012, adamo_high-resolution_2013, claeyssens_star_2023}. We therefore compared the ``lil gal'' SE stitched continuum to stellar populations at ages ranging from the upper limit of 1 Gyr (reddened by E(B-V)=0.25) to older populations of 5 Gyr, applying a redshift of 0.9225. Stitching the continuum based on the stellar populations involved applying correction factors to the flux so that the overlapping edges of the orders matched. Order 5 was shifted up by 11\% and order 3 was shifted down by 14\%. We find that both the 1 Gyr and 5 Gyr models yield reasonable fits, so we cannot place strong constraints on the age of the SE component, nor confirm or rule out the clumpy disk interpretation based on age alone. However, we do confirm that a redshift of 0.9225 yields a consistent match with stellar population models, supporting the conclusion that the SE continuum is of stellar origin at that redshift.

To probe the conditions of the ionized gas, we use the \sii~$\lambda$6716/$\lambda$6731 line ratio as a diagnostic of electron density. However, note that we only detect lines corresponding to $z$ = 0.9248 at 2.2$\sigma$ and 1.4$\sigma$ significance from the total extraction --- so the result should be interpreted with caution. Furthermore, while the NW component can be identified in the total-system extraction, the lines are less apparent in the dedicated NW extraction (see middle panel of Figure \ref{fig:linefit_combined}). The measured ratio of $\sim$1.6 from the integrated spectrum exceeds the theoretical upper limit of 1.4 from \cite{osterbrock_astrophysics_2006}. \citet{krabbe_interaction_2014} report similar ratios in H\textsc{ii} regions of interacting galaxies, suggesting that such deviations can occur in complex environments. Other studies, including \citet{kennicutt_comparison_1989} and \citet{lopez-hernandez_integral_2013}, have also reported \sii~$\lambda$6716/$\lambda$6731 ratios above the expected theoretical maximum. \citet{lopez-hernandez_integral_2013} further notes that ratios $\gtrsim$\,1.4 typically indicate electron densities below 10 cm$^{-3}$, and precise values become increasingly uncertain as the ratio approaches the asymptotic limit. In such cases, a density of 100 cm$^{-3}$ is often assumed.

Building on the NIRES analysis, KCWI observations reveal additional spectroscopic features of the system. Given the pixel scale of 0.\arcsec29 px$^{-1}$ along the slicer, 0.\arcsec68 px$^{-1}$ perpendicular to the slicer, and the angular separation of 0.\arcsec7 between the components, we cannot spatially resolve the NW companion from the underlying galaxy as was done with the NIRES data. Furthermore, given the red spectral slope of the NW source, it is difficult to distinguish the exact location of the rest-frame optical emission seen by KCWI.
We highlight the H$\beta$, \oiii~4959 \text{\AA}, and \oiii~5007 \text{\AA} emission lines from the KCWI extraction plotted in Figure~\ref{fig:kcwispectra}. Unlike in the NIRES data, where the two components of the close pair system at $z = 0.9225$ and $z = 0.9248$ are distinctly resolved, the KCWI spectrum shows a blending effect, with a single set of emission lines centered at an intermediate redshift of $z \sim0.9235$. We note that the \oiii~4959 \text{\AA} line in Figure~\ref{fig:kcwispectra}, expected to be 1:2.98 of the \oiii~5007 \text{\AA} line \citep{storey_theoretical_2000}, is likely buried under the prominent skylines that dominate this region of the spectrum. We treat the extracted KCWI spectrum as representing emission from the total system in the ensuing line flux calculations.

From the Gaussian fiting results reported in Table \ref{tab:linefit_results}, we calculate the log(\nii~6583 \text{\AA}/\ha) line flux ratios for the total system, NW, and SE components: $-0.41 \pm 0.14, -0.49 \pm 0.22, -0.21 \pm 0.09$, respectively. Since we cannot spectrally resolve the two components in the KCWI data, we employed a simpler three-component compound Gaussian model in Figure \ref{fig:kcwispectra} to fit the \hb~(red), \oiii~4959 \text{\AA} (blue), and \oiii~5007 \text{\AA} (green) lines. The resulting KCWI line flux ratio for log(\oiii~5007 \text{\AA})/\hb) is -0.35 \(\pm\) 0.17.

Using the \ha~and \hb~emission line fluxes and equation 7 of \cite{dominguez_dust_2013}, we estimate the dust extinction $A_{V}$. This yields a value of $A_{V} = 10.22 \pm 0.27$ mag, consistent with the high obscuration typically observed in the nuclear regions of starburst galaxies \citep{mattila_highly_2004}, where $A_{V}$ can reach up to $\sim$ 25$-$40 mag in merging \added{(U-)LIRGS} \citep{mattila_adaptive_2007,vaisanen_shutting_2017,u_keck_2019}.

\begin{deluxetable*}{lcccc}[ht!]
\tablecaption{\label{tab:linefit_results} Emission Line Gaussian Fitting Results}
\tablehead{
    \colhead{Line} & \colhead{Component} & \colhead{Flux ($10^{-16}\ \mathrm{erg\,s^{-1}\,\text{\AA}^{-1}\,cm^{-2}}$)}
 & \colhead{Mean $\lambda_{observed}$(\text{\AA})} & \colhead{$\sigma$ (\text{\AA})}
}
\startdata
\hb & Total System & 4.71 $\pm 0.67$ & 9350.28 $\pm 0.46$ & 1.70 $\pm 0.46$\\
\oiii~4959 \text{\AA} & Total System & 0.71 $\pm 0.27$ & 9537.76 $\pm 5.62$ & 2.52 $\pm 1.89$\\
\oiii~5007 \text{\AA} & Total System& 2.11 $\pm 0.92$& 9631.35 $\pm 1.89$ & 2.52 $\pm 1.89$\\
\nii~6548 \text{\AA} & Total System & 33.15 $\pm 6.52$ & 6548.97 $\pm 3.83$ & 3.02 $\pm 1.26$\\
\nii~6548 \text{\AA} & NW Narrow & 14.71 $\pm 2.39$ & 6547.91 $\pm 0.54$ & 1.36 $\pm 0.55$\\
\nii~6548 \text{\AA} & NW Broad & 1.86 $\pm 0.07$ & 6547.63 $\pm 3.72$ & 3.89 $\pm 3.69$\\
\nii~6548 \text{\AA} & SE & 11.64 $\pm 2.64$ & 6542.97 $\pm 0.20$& 2.39 $\pm 1.22$\\
\ha & Total System & 257.65 $\pm 15.26$ & 6562.67 $\pm 0.46$ & 3.19 $\pm 0.53$\\
\ha & NW Narrow & 96.98 $\pm 4.87$ & 6563.09 $\pm 0.11$ & 1.35 $\pm 0.17$\\
\ha & NW Broad & 58.0 $\pm 12.02$ & 6562.53 $\pm 1.14$& 3.89 $\pm 2.01$\\
\ha & SE & 39.2 $\pm 8.05$ & 6555.19 $\pm 0.81$ & 1.59 $\pm 0.82$\\
\nii~6583 \text{\AA} & Total System & 99.47 $\pm 3.53$ & 6583.51 $\pm 0.69$& 3.31 $\pm 0.24$\\
\nii~6583 \text{\AA} & NW Narrow & 44.12 $\pm 3.64$ & 6583.55 $\pm 0.25$ & 1.36 $\pm 0.28$\\
\nii~6583 \text{\AA} & NW Broad & 5.58 $\pm 0.69$ & 6581.57 $\pm 0.28$ & 3.89 $\pm 0.37$\\
\nii~6583 \text{\AA} & SE & 24.33 $\pm 2.78$ & 6575.18 $\pm 0.24$ & 1.66 $\pm 0.47$\\
\enddata
\end{deluxetable*}

\subsection{BPT Classification}\label{bpt}
\begin{figure}[h!]
\centering
\includegraphics[width=0.5\textwidth]{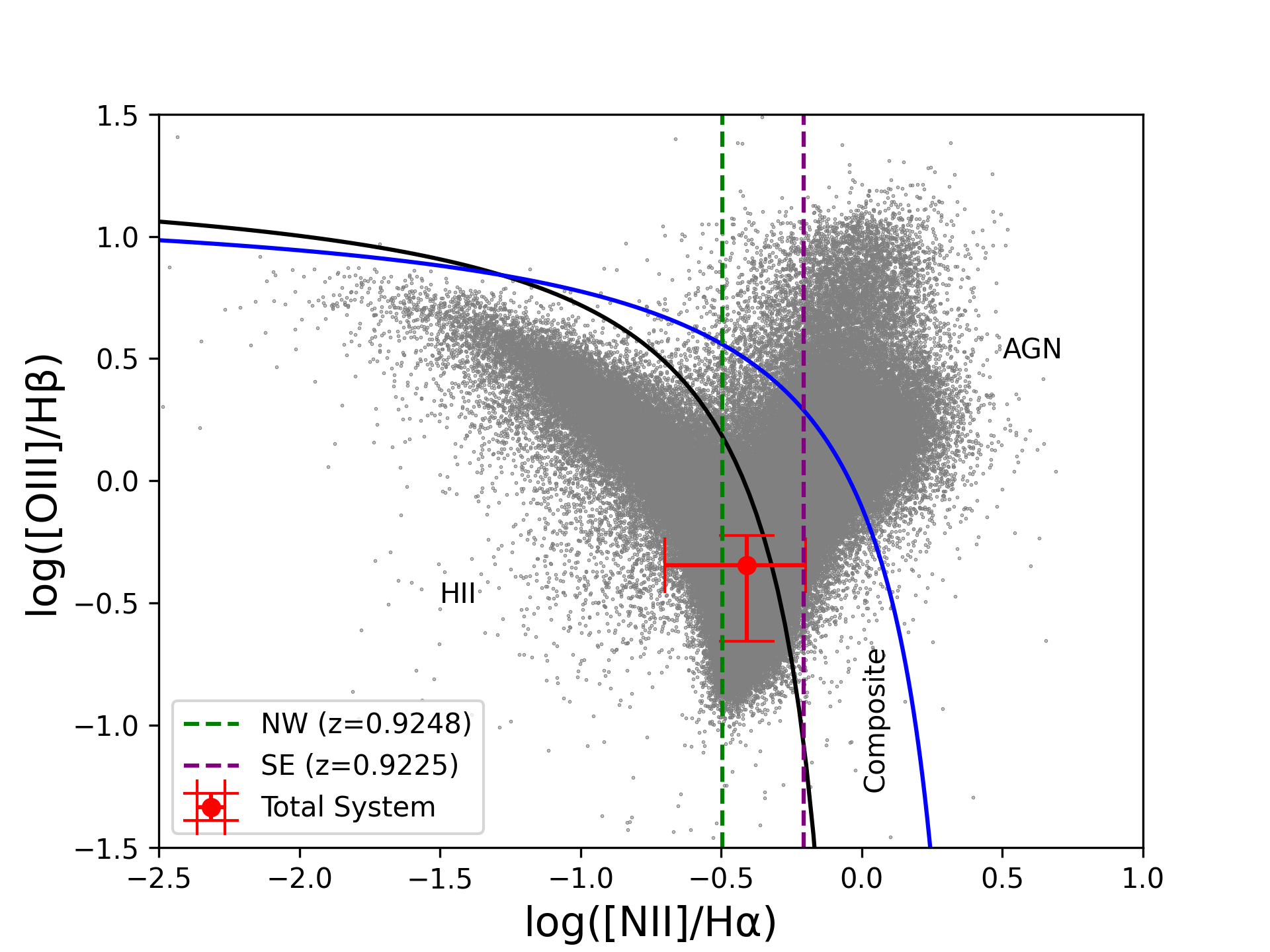}
\caption{\label{fig:bpt}BPT diagram of log(\nii~6583 \text{\AA}/\ha) and log(\oiii~5007 \text{\AA}/\hb) of the system (red), NW component (green), and SE component (purple), determined from NIRES and KCWI spectra. The theoretical classification bounds define the AGN (blue) and H\textsc{ii} SF (black) ionization mechanisms \citep{kewley_theoretical_2001,kauffmann_host_2003}, plotted with the Sloan Digital Sky Survey (SDSS) Data Release 16 (gray). The NW companion and total system fall within the H\textsc{ii} region boundary, indicating active star formation. The SE component lies within the composite region.}
\end{figure}

Further spectroscopic analysis can provide insights into the photoionization mechanisms of the galaxy system. Central to this analysis are BPT/VO diagrams, which correlate specific emission line ratios, such as log(\nii~6583 \text{\AA}/\ha) and log(\oiii~5007 \text{\AA}/\hb), to discern different ionization mechanisms within galaxies \citep{baldwin_classification_1981, veilleux_spectral_1987}. The diagram distinguishes active star formation from AGN ionization mechanisms based on theoretical models dependent on these emission line ratios, detailed in \cite{kewley_theoretical_2001} and \cite{kauffmann_host_2003}. These models are plotted as the blue and black curves in Figure \ref{fig:bpt}. 

We plot the diagnostic ratios on the BPT diagram in Figure \ref{fig:bpt}, using vertical lines to indicate the NW (green) and SE (purple) components, since we cannot calculate their respective log(\oiii~5007 \text{\AA})/\hb) values from the KCWI data. The total system and NW component lie within the H\textsc{ii} region of the BPT, indicating active star formation. This is consistent with starburst galaxies triggered by mergers at $z$ \(\sim\) 1 \citep{diaz-santos_herschelpacs_2017, renaud_mergerstarburst_2022}. The SE component appears to lie within the composite classification, falling under the upper limit of the theoretical starburst model from \cite{kewley_theoretical_2001}. \cite{yuan_role_2010} report that starburst and starburst-AGN composite galaxies dominate early merger stages at nuclear separations $< 10$ kpc, providing us with further insights into the nature of this merging system. 
The position of ``lil gal'' on the BPT is also consistent with \cite{kewley_host_2006}'s classification of star-forming galaxy pairs ($s < 20$ kpc $h^{-1}$), lying closer to the classification line than isolated galaxies. 
\cite{rich_galaxy_2015} found that composite BPT positions in local merging LIRGs are commonly produced by merger-driven shocks rather than AGN, with the shock contribution emerging at close-pair separations ($<$ 10 kpc). \cite{newman_nebular_2013} showed that composite positions in $z \sim 2$ star-forming galaxies often reflect elevated ISM conditions or shocks rather than AGN activity. At a projected separation of 5 kpc, the different \nii/\ha~ratios of the NW and SE components are consistent with this picture.
Although we were unable to resolve the two components in the KCWI spectra --- due primarily to the spatial resolution of the seeing-limited KCWI observations but perhaps also to the red slope of the NW source, these results highlight the critical role of follow-up spectroscopic observations in probing the photoionization structure and merging nature of the galaxy system. 

\subsubsection{Metallicity Estimates}\label{metallicity}
\begin{deluxetable}{lc}[h!]
\tabletypesize{\small}
\tablecaption{Metallicity Estimates\label{tab:metallicity}}
\tablehead{
\colhead{Method} & \colhead{12 + log(O/H)}
}
\startdata
(1) $12 + \log(\mathrm{O/H}) = 8.9 + 0.57\times N2$ & 8.67 $\pm$ 0.057 \\
(2) $12 + \log(\mathrm{O/H}) = 8.73 - 0.32\times O3N2$ & 8.72 $\pm$ 0.107 \\
(3) $12 + \log(\mathrm{O/H}) = 9.12 + 0.73\times N2$ & 8.82 $\pm$ 0.073 \\
(4) $\mathrm{O/H} = 4.15 \times 10^{-4}x_{3}^{-0.29}x_{10}^{0.40}$  & 8.90 $\pm$ 0.048 \\
\enddata
\tablenotetext{1}{\scriptsize $N2 = \log($[\ion{N}{2}]~6583~\AA/\ha) from \cite{pettini_o_2004} (PP04 hereafter).}
\tablenotetext{2}{\scriptsize $O3N2 = \log(($[\ion{O}{3}]~5007~\AA/\hb)/([\ion{N}{2}]~6583~\AA/\ha)) from PP04.}
\tablenotetext{3}{\scriptsize From \cite{denicolo_new_2002}.}
\tablenotetext{4}{\scriptsize  $x_3 =$ [\ion{O}{3}]~5007~\AA/\hb, $x_{10} =$ [\ion{N}{2}]~6583~\AA/\ha~from \cite{charlot_nebular_2001}.}

\end{deluxetable}
Across the various methods to estimate metallicity reported in Table \ref{tab:metallicity}, we find that the ``lil gal''system has metallicities near the solar value. \cite{stott_relationship_2014} report a median metallicity of $8.63 \pm 0.11$ using Method 1 for $z \sim 1$ star-forming galaxies, consistent with our results using the same methods. \cite{marquez_rotation_2002} presents a trend between the \nii~6583 \text{\AA}/\ha~metallicity indicator (Method 3) and the morphological type and interaction status. Comparing our ratios, we find that ``lil gal'' (\nii~6583 \text{\AA}/\ha~= 0.4) is consistent with their classification criteria for interacting early-type spiral galaxies (\nii~6583 \text{\AA}/\ha~= 0.38). In terms of the luminosity-metallicity (L--Z) relation, we expect mergers to exhibit depressed metallicities below the L--Z relation \added{\citep{2008AJ....135.1877E, rupke_oxygen_2008}}. We therefore compare the metallicity of ``lil gal'' to the expected metallicity at $z \sim 1$ for a galaxy of the luminosity we reported in \S\ref{photom}. \cite{lamareille_spectrophotometric_2006} derives an L--Z relation for $z = 0.2–1$ galaxies using oxygen abundances estimated with Method 4 and B-band magnitudes. The 8.9 abundance value for ``lil gal'' falls below the expected value of 9.24 from the \cite{lamareille_spectrophotometric_2006} L--Z relation at ``lil gal's'' B-band magnitude of -21.1, which follows the expected trend mergers falling below the L--Z relation. Furthermore, the 8.9 abundance value report for ``lil gal'' agrees with the \cite{kewley_metallicity_2006} L--Z relation for interacting pairs with similar separations and B-band magnitudes.

\section{Summary and Conclusion}

\added{The case of ``lil gal'' illustrates an interpretive challenge for multi-band \textit{JWST} imaging: a source that appears as an isolated disk at 5.6 and 7.7 $\mu$m but reveals a secondary peak at 15 $\mu$m. We evaluated three potential scenarios to explain this wavelength-dependent morphology: (1) a clumpy star-forming disk with a dust-obscured region, (2) a chance superposition of an interloper, or (3) a physical merger at $\sim$5 kpc separation. By synthesizing high-resolution morphology, multi-band SED fitting, integrated and resolved spectroscopy, and spatial profile analysis, we have conclusively shown that this system is a physical merger. This case study establishes a robust, multi-wavelength verification framework for classifying similar ambiguous, red-excess sources being uncovered in current and future \textit{JWST} imaging surveys.

\subsection*{Morphological Evidence:}
The morphological observations and GALFIT results from the MIRI data --- including the distinct emission from both components at 15 $\mu$m, low Sérsic indices, and rising $\chi^2$ residuals with wavelength --- are consistent with a disturbed, non-isolated system, with morphologies resembling those seen in confirmed mergers \citep{margalef-bentabol_formation_2016, he_close_2022}. However, morphology alone cannot distinguish between these scenarios at the spatial resolution of MIRI at $z \sim 1$, as a clumpy disk with a dust-obscured off-center region could produce a similar two-component structure. This limitation is consistent with the known challenges of optical merger classification at this epoch \citep{hung_comparison_2014}. The ``lil gal'' system, barely detected in \textit{HST} imaging and only exhibiting its dual-peaked structure at 15 $\mu$m, underscores the need for additional diagnostics to identify mergers that morphological classification alone would miss.
\subsection*{Photometric Evidence:}
The photometric redshift of $z \sim 0.92$ places the system at an epoch where merger rates are expected to be elevated, though precise fractions remain uncertain due to the methodological challenges \citep{kaviraj_galaxy_2015, duncan_observational_2019}. The infrared hump in the SED is characteristic of dusty starburst galaxies, a signature often found in merging systems. Furthermore, the alignment between the photometric data and the SED models of peculiar, merging systems supports the merger scenario for the ``lil gal'' system. Photometric redshift alone, however, cannot rule out a chance-superposition interloper: integrated SED fitting cannot distinguish whether the two morphological components share a common redshift or are projected at different distances.

\subsection*{Kinematic Evidence:}
The Keck NIRES data resolve \ha, \nii~$\lambda\lambda$ 6548,6583 \text{\AA}, and He I 10830 \text{\AA} at $z = 0.9248$ in the NW component and at $z = 0.9225$ in the SE component, with the observed velocity offset of 360 km s\(^{-1}\) consistent across multiple emission lines. The close but distinct redshifts rule out the chance-superposition hypothesis. The 360 km s$^{-1}$ offset further falls within the $\Delta v \leq 500$ km s$^{-1}$ criterion for gravitationally bound close pairs, lending strong kinematic evidence that the NW and SE sources are interacting components of a galaxy merger system. 

\subsection*{Spatial Profile Analysis:}
Distinguishing the merger and clumpy-disk scenarios requires verifying that the two velocity components correspond to spatially distinct emission regions. The spatial profile analysis in Figure \ref{fig:spatial_profiles} shows that the emission peaks at $z = 0.9248$ and $z = 0.9225$ are offset along the slit direction across three emission lines, confirming the two-component nature of the system. 

\subsection*{Diagnostic Emission Line Ratios:}
The elevated SFR of the ``lil gal'' system supports a merger-driven scenario. Placed within the H{\sc ii} region of the BPT diagram, the system exhibits characteristics of a starburst galaxy, often associated with interactions at $z \sim 1$ \citep{diaz-santos_herschelpacs_2017, renaud_mergerstarburst_2022}. The SE component's composite classification is consistent with merger-driven excitation rather than AGN activity at this separation (\S\ref{bpt}). In addition, the metallicity of the system falls below the L--Z relation at $z \sim 1$ \citep{lamareille_spectrophotometric_2006}, offering additional spectroscopic support for the merger hypothesis.

\bigskip

Taken together, this multi-wavelength approach --- combining MIRI morphology, SED constraints, and ground-based spatially-resolved spectroscopy --- provides a template for classifying ambiguous sources in \textit{JWST} surveys. The appearance of ``lil gal'' as a single disk at shorter wavelengths, with the secondary component emerging only at 15 $\mu$m, illustrates that coordinated imaging and spectroscopy can resolve interpretive ambiguities that single-band morphology cannot, providing a path forward for similar sources at cosmic noon and beyond.}

\facility{JWST (MIRI), HST (ACS, WFC3), Keck:II (NIRES, KCWI)}

\software{Astropy~\citep{astropy_collaboration_astropy_2013, astropy_collaboration_astropy_2018, astropy_collaboration_astropy_2022}, GALFIT~\citep{peng_detailed_2002,peng_detailed_2010},
Matplotlib~\citep{hunter_matplotlib_2007},
PHOTUTILS~\citep{bradley_astropyphotutils_2023},
Scipy~\citep{virtanen_scipy_2020},
SPECUTILS~\citep{earl_astropyspecutils_2024}}

\section*{Acknowledgements}
\added{The authors are grateful to the anonymous referee for their constructive feedback and insightful suggestions, which significantly improved the clarity and presentation of this manuscript.}
We acknowledge helpful discussions with Matt Malkan, Thomas Lai, Lee Armus, and the rest of the GOALS Team on the spectral analysis part of the project. This work is based in part on observations made with the NASA/ESA/CSA James Webb Space Telescope and the NASA/ESA Hubble Space Telescope. The data were obtained from the Mikulski Archive for Space Telescopes at the Space Telescope Science Institute, which is operated by the Association of Universities for Research in Astronomy, Inc., under NASA contracts NAS 5-03127 for JWST and NAS 5-26555 for HST. The specific observations analyzed can be accessed via: \dataset[https://doi.org/10.17909/1cnz-fd50]{https://doi.org/10.17909/1cnz-fd50}. These observations are associated with program \#JWST-GO-01717 and HST PIDs \#16914, 10592, and 14095.  
Some of the data presented herein were obtained at Keck Observatory, which is a private 501(c)3 non-profit organization operated as a scientific partnership among the California Institute of Technology, the University of California, and the National Aeronautics and Space Administration. The Observatory was made possible by the generous financial support of the W. M. Keck Foundation.
The authors wish to recognize and acknowledge the very significant cultural role and reverence that the summit of Maunakea has always had within the Native Hawaiian community. We are most fortunate to have the opportunity to conduct observations from this mountain.

V.U. gratefully acknowledges partial funding support from NASA ADAP grant \#80NSSC23K0750 and ADSPS grant \#80NSSC25K0169, National Science Foundation (NSF) Astronomy and Astrophysics Research Grant \#2536603, as well as \emph{STScI} grants \#JWST-GO-08391.001-A, \#JWST-GO-01717.001-A, \#HST-AR-17063.005-A, and \#HST-GO-17285.001-A, which were provided by NASA through a grant from the Space Telescope Science Institute, which is operated by the Association of Universities for Research in Astronomy, Inc., under NASA contract NAS 5-03127. V.U and G.C. had partially performed work for this project at the Aspen Center for Physics, which is supported by National Science Foundation grant PHY-2210452. T.G. acknowledges support from the Australian Research Council through Discovery Project DP210101945, funded by the Australian Government.

\bibliography{thesis_ref}{}
\bibliographystyle{aasjournalv7}

\end{document}